\documentclass[aps,pre,superscriptaddress,floatfix,12pt]{revtex4-1}

\usepackage{amsmath,amsfonts,amssymb}
\usepackage{graphicx, tikz}
\usepackage{wrapfig}
\usepackage{algorithm}
\usepackage{algorithmic}
\usepackage{epstopdf}
\usepackage{hyperref}
\usepackage{caption}
\usepackage{subcaption}
\usepackage{mathbbol,bbm}

\global\long\def\slfrac#1#2{\left.#1\middle/#2\right.}

\let\originalleft\left
\let\originalright\right
\renewcommand{\left}{\mathopen{}\mathclose\bgroup\originalleft}
\renewcommand{\right}{\aftergroup\egroup\originalright}

\newcommand{\ie}{\emph{i.e.}}

\newcommand{\etal}{{\it et al. }}

\hypersetup{
	pdfproducer={}  
}

\begin{document}
\title{Fast inference of ill-posed problems within a convex space}

\author{J. Fernandez-de-Cossio-Diaz} 
\email{cossio@cim.sld.cu}
\affiliation{Department of Systems Biology, Center of Molecular Immunology, La Habana, Cuba}

\author{R. Mulet}
\email{mulet@fisica.uh.cu}
\affiliation{Group of Complex Systems and Statistical Physics. Department of Theoretical Physics, Physics Faculty, University of Havana, Cuba}

\date{\today}

\begin{abstract}
In multiple scientific and technological applications we face the problem of having low dimensional data to be justified by a linear model defined in a high dimensional parameter space. The difference in dimensionality makes the problem ill-defined: the model is consistent with the data for many values of its parameters. The objective is to find the probability distribution of parameter values consistent with the data, a problem that can be cast as the exploration of a high dimensional convex polytope. In this work we introduce a novel algorithm to solve this problem efficiently. It provides results that are statistically indistinguishable from currently used numerical techniques while its running time scales linearly with the system size. We show that the algorithm performs robustly in many abstract and practical applications. As working examples we simulate the effects of restricting reaction fluxes on the space of feasible phenotypes of a {\em genome} scale E. Coli metabolic network and infer the traffic flow between origin and destination nodes in a real communication network.
\end{abstract}

\maketitle
\section{Introduction}

Although we live in the time of high-throughput data, or perhaps because of it, today many problems in science and technology have to deal with experiments that provide low dimensional data and a huge parameter space to be explored in search for a comprehension of this data. For example, in Metabolic Network Analysis \cite{schellenberger2009use_5457} a stoichiometric matrix of $N$ reactions between $M$ metabolites is given, and one is interested in estimating the fluxes through these reactions. In Network Tomography \cite{vardi1996network_365} data flows on $M$ peer-to-peer router links are known and the interest is to estimate the net flow of data between $N$ pairs of origin to destination nodes. In both cases $N$ is larger than $M$. Similar problems arise in other fields like Compressed Sensing \cite{krzakala2011statistical_}, Image Super-resolution \cite{shepp1978computerized_420}, Positron Emission Tomography \cite{vardi1985statistical_8}, the freezing transition of hard spheres \cite{kapfer2013sampling_1301} and density reconstruction from gravitational lensing in astrophysics \cite{lubini2012sampling_3077}. All these examples may be recognized as linear ill-posed inverse problems where the dimension of the data available is smaller than the number of parameters to be explored looking for a consistent model. The problem is usually cast as the set of linear equations: $\vec{y} = {\bf S} \vec{x}$ with $\vec{x} \in [\vec{a},\vec{b}]$, where the dimension of $\vec{y}$ is lower than the dimension of $\vec{x}$. Given $\vec{y}$ and ${\bf S}$ the task is to make inferences about $\vec{x}$.

In specific contexts it is known that the vector $\vec{x}$ is sparse, where the sparseness is defined by a properly chosen (sometimes arbitrarily) distance. In this case the problem translates into an optimization problem: finding the sparsest vector in the set of possible solutions \cite{donoho2003optimally_2197, candes2006robust_48}.  Alternatively, the linear set of equations may be complemented with a linear objective function to be optimized. This approach has been very successful in Metabolic Network Analysis \cite{palsson2006systems} where the function to optimize has a reasonable biological meaning, for example, the growth rate of the cell, and can be maximized by Linear Programming techniques \cite{papadimitriou1998combinatorial}. However, it seems difficult to justify similar strategies in other fields, where the sparseness is not a reasonable ansatz, or where it is impossible to have an intuition about which function to maximize or minimize.

A more general approach is to introduce a Maximum Entropy-like principle to be satisfied by the data, usually with the addition of some sort of side information that, depending on its pertinence or quality, may be good or bad for the inference. In this case the solution will depend, not only on the technique used to infer $\vec{x}$, \ie  Entropy Maximization (EM), Maximum Likelihood (ML), Markov Random Fields, Hierarchical Clustering, Vector Auto-regressive Models (VAR) etc., \cite{werhli2012comparing_52, beerli2006comparison_341, frey2005comparison_1, kolaczyk2009statistical} but also on the prior proposed. For example, in the context of Network Tomography, Medina \etal showed that standard inference methods in this field were sensitive to prior assumptions \cite{medina2002traffic_}, and similar conclusions were obtained inferring gene interaction networks \cite{olsen2014_177}, studying soil water dynamics \cite{scharnagl2011_3043}, gene selection \cite{verbyla2010sensitivity_} and medical imaging \cite{lalush1992simulation_267}, just to cite a few examples.

Within these approaches the ill-defined linear problem is transformed into a well defined one, with {\em one} solution relatively easy to find. The price to pay is the introduction of ad-hoc assumptions that may be unjustified or may introduce unwanted biases hard to control. Without these assumptions it is mandatory to infer {\em all} the solutions of the system $\vec{y} = {\bf S}\vec{x}$, more specifically the probability of occurrence of each solution. This amounts to finding the points inside the convex polytope defined by a linear set of equations, or equivalently the volume of the high dimensional convex body \cite{dyer1988complexity_967,beuler2000polytopes_131}.

The exact determination of this volume is known to be $\#P$-{\em hard} and therefore exact algebraic methods are computationally infeasible even in low dimensional problems \cite{dyer1988complexity_967}. On the other hand stochastic approaches, like  the {\em hit and run} Markov Chain Monte Carlo (MCMC) \cite{smith1984hit_1296, turcin1971computation_720, demartino2014uniform_1312} are widely used but still too expensive for very high dimensional problems. A different stochastic algorithm, based on message passing techniques borrowed from the information theory community, was recently introduced in \cite{braunstein2008estimating_240} to study the volume of the phenotypic space of Metabolic Networks, and latter extended in \cite{massucci2013novel_838, font-clos2012weighted_11003}. Unfortunately, the good performance and  efficiency of message passing techniques in problems with binary variables \cite{mezard2009information,KS,PREcol} is limited here because of the {\em continuous} and {\em bounded} support of the quantities involved.

To overcome this limitation we  introduce a proper parametrization of the messages. The idea has been borrowed from the study of message passing equations with continuous variables \cite{mezard2009information,krzakala2011statistical_}, but its application in problems with bounded support has been elusive for years. The reason is that in this case one must consider, together with the moments of the distributions, their corresponding bounds.  We present in this work the solution to this problem as a set of fixed point equations and its corresponding implementation as a message passing algorithm over a finite number of parameters. 

We test the algorithm performance on different ensembles of artificial graphs and show that it has very good convergence properties, provides  results that are statistically equivalent to those given by {\em hit and run} Markov Chain Monte Carlo (MCMC), while it is much faster - it scales linearly with system size. To further prove the capabilities of our algorithm we explore the effects of reaction knock-down experiments in a genome scale E. Coli metabolic network, unveiling the importance of redundancy in the network. Finally, we use our algorithm to infer the traffic flow between origin and destination nodes in the Abilene Network of the USA and show that our predictions compare remarkably well with the experimental available data.

\section{Methods}
\label{sec:met}

We want to study the space of feasible solutions $\vec{x}$ subject to the constraints:
\begin{equation}
\vec{y} = {\bf S}\vec{x},\qquad
\vec{a}\le\vec{x}\le\vec{b}.
\label{eq:matrix},
\end{equation}
where ${\bf S}$ is an $M\times N$ matrix, $M < N$. Equation \eqref{eq:matrix} defines a convex polytope in a ($N-M$)-dimensional manifold embedded in the $N$-dimensional space of the variables (we assume that {\bf S} is full rank). To simplify the notation for the rest of this section we will set $\vec{y}=0$ without loss of generality (it is always possible to translate the origin of coordinates with an arbitrary solution of $\vec{y} = {\bf S}\vec{x}$). The solution of \eqref{eq:matrix} is not unique, and in the absence of further information, we must introduce a uniform probability distribution over the solutions:
\begin{equation}
\mathrm{P}(\vec{x}) \propto \delta({\bf S}\vec{x})
\label{eq:probframe}
\end{equation}
for $\vec{a}\le\vec{x}\le\vec{b}$, and $\mathrm{P}(\vec{x}) = 0$ otherwise, where $\delta(\vec{x})$ is the multi-dimensional Dirac delta function and the proportionality factor is a normalization constant. The Bethe approximation is the simplest approximation to the problem and assumes that the factor graph of constraints is a tree. In this case:
\begin{equation}
\mathrm{P}(\vec{x}) =	
	\prod_a \mathrm{P}_a (\{x_j\}_{j \in a}) \prod_i \mathrm{P}_i (x_i)^{1-d_i},
\label{bethe}
\end{equation}
where $i$ is a variable (node) index, $a$ is an equation (factor node) index, $d_i$ is the number of equations in which the variable $x_i$ participates (i.e. the degree of site $i$) and 
\begin{align}
\mathrm{P}_a(\{x_j\}_{j \in a}) & =
	\int \mathrm{P}(\vec{x}) \mathrm{d} \{x_j\}_{j \notin a}
	\label{eq:Pa} \\
\mathrm{P}_i(x_i) & =
	 \int \mathrm{P}(\vec{x}) \mathrm{d} \{x_j\}_{j \neq i}
	\label{eq:Pi}
\end{align}
represent marginal probability distributions. Of course one may wonder whether this approximation is useful or not in a real-world situation. It turns out that, in fact, it has worked successfully in a number of applications, for instance, constraint satisfaction problems \cite{mezard2009information}, error correcting codes \cite{richardson2001capacity_599}, perceptron learning \cite{baldassi2007efficient_11079} and metabolic networks \cite{braunstein2008estimating_240}. The intuition is that if typical loop lengths are large enough, a tree is a good approximation to the statistical correlations in the network.

Within this approximation the problem can be reduced to the solution of an iterative set of equations, usually called Belief Propagation algorithm, for two kind of messages, $m_{a\rightarrow i}(x_i)$ from the equation (factor node) $a$ to the variable (node) $x_i$ and $n_{j\rightarrow a}(x_j)$ from the variable $x_j$ to the factor node $a$.
\begin{align}
m_{a\rightarrow i}(x_i) & \propto
	\int_{\{x_j\}_{j \in a \backslash i}} 
	\delta\left(\sum_{i \in a} S_{ai} x_i\right)
	\prod_{j\in a\backslash i} n_{j\rightarrow a}(x_j)
	\label{eq:update_ai} \\
n_{i\rightarrow a}(x_i) & \propto
	\prod_{c\in i\backslash a} m_{c\rightarrow i}(x_i)
	\label{eq:update_ia}
\end{align}
In practical terms this iteration is particularly cumbersome in the current problem of interest because $x_i$ is defined on a continuous and bounded support\cite{braunstein2008estimating_240}. In fact, although the number of message passing equations scales linearly with the system size, the continuos support imposes convolution and multiplication operations, whose costs scale as $D^{d_a-1}$, where $D$ is the discretization chosen for the distribution and $d_a$ is the number of variables connected to a factor node. 

An efficient solution is to parameterize the messages and to write the update rules \eqref{eq:update_ai} and \eqref{eq:update_ia} in terms of these parameters, significantly reducing the dimensionality of the stored arrays of values and the computations per iteration. In a convex body, we know that the distributions will be unimodal, and in general non-symmetric in the vicinity of the mode. A simple distribution satisfying these conditions is a generalized Beta distribution, suggesting that:
\begin{equation}
m_{a\rightarrow i}(x_i)
\propto(x_i-A_i)^{\alpha_i-1} (B_i-x_i)^{\beta_i-1}
\end{equation}
This is a proper parametrization of the messages, with the constrains $\alpha_i,\beta_i\ge 1$ and $a_i \le A_i \le B_i \le b_i$. The implementation of Belief Propagation that follows from this parameterization will be called BP$\beta$ in the rest of this paper. In the Supplementary Material we show how the iterative scheme \eqref{eq:update_ai}-\eqref{eq:update_ia} can be approximated by update equations consistent with this parameterization. In short, the message passing equations for $\mu_{a\rightarrow i}$ and $\sigma_{a\rightarrow i}$, the mean and variance of the Beta distributions, read:
\begin{equation}
\mu_{a\rightarrow i} = \sum_{j\in a\backslash i}
	\frac{S_{aj}}{S_{ai}} \mu_{j\rightarrow a}, \quad
\sigma_{a\rightarrow i}^2 = \sum_{j\in a\backslash i}
	\frac{S_{aj}^2}{S_{ai}^2} \sigma_{j\rightarrow a}^2
\end{equation}
The updating rules for $A_{a\rightarrow i}$ and $B_{a\rightarrow i}$, the lower and upper bounds of the messages, are:
\begin{equation}\begin{aligned}
A_{a\rightarrow i} & =
	- \sum_{j\in a^- \backslash i} \frac{S_{aj}}{S_{ai}} A_{j\rightarrow a}
	- \sum_{j\in a^+ \backslash i} \frac{S_{aj}}{S_{ai}} B_{j\rightarrow a}
\\
B_{a\rightarrow i} & =
	- \sum_{j\in a^- \backslash i} \frac{S_{aj}}{S_{ai}} B_{j\rightarrow a}
	- \sum_{j\in a^+ \backslash i} \frac{S_{aj}}{S_{ai}} A_{j\rightarrow a}
\end{aligned}
\end{equation}
where $j \in a^+$ denotes the set of $j$'s for which $\slfrac{S_{aj}}{S_{ai}}>0$, and $j\in a^-$ denotes those for which $\slfrac{S_{aj}}{S_{ai}}<0$.  From $\mu_{a\rightarrow i}$, $\sigma_{a\rightarrow i}^2$, $A_{a\rightarrow i}$ and  $B_{a\rightarrow i}$ we can obtain closed form equations for the parameters $\alpha_{a\rightarrow i}$ and $\beta_{a\rightarrow i}$. The equations for the messages from variables to nodes $n_{i\rightarrow a}$ are readily obtained by single variable numerical integrations. The details of this derivation are given in the Supplementary Materials.

One the other hand, it is possible to define an entropy in terms of the marginal probability distributions. This amounts to the logarithm of the volume of the convex solution space. Still assuming that the factor graph is a tree, one obtains the following expression:
\begin{equation}
H = \sum_a H_a - \sum_i (d_i - 1) H_i,
\label{eq:entropy}
\end{equation}
where $H_a$ is the joint entropy of the variables participating in equation $a$, and $H_i$ is the entropy of variable $i$,
\begin{align}
H_a & = \ln \int f_a(\vec{x}_a) \prod_{i\in a} n_{i\rightarrow a} (x_i) \mathrm{d}\vec{x}_a - \sum_{\langle ai \rangle} \int b_i(x_i) \ln n_{i\rightarrow a}(x_i) \mathrm{d}x_i \label{eq:Sa} \\
H_i & = \int b_i(x_i) \ln b_i(x_i) \mathrm{d}x_i \label{eq:Si}
\end{align}
The entropy is calculated once the belief propagation iteration has converged (see also the Supplementary Materials for more details.)

\section{Results}
\label{sec:res}

\subsection{Statistical Analysis}
\label{subsec:statana}

\begin{figure}
\centering
	\begin{subfigure}{.45\textwidth}
		\centering
		\includegraphics[width=\linewidth]{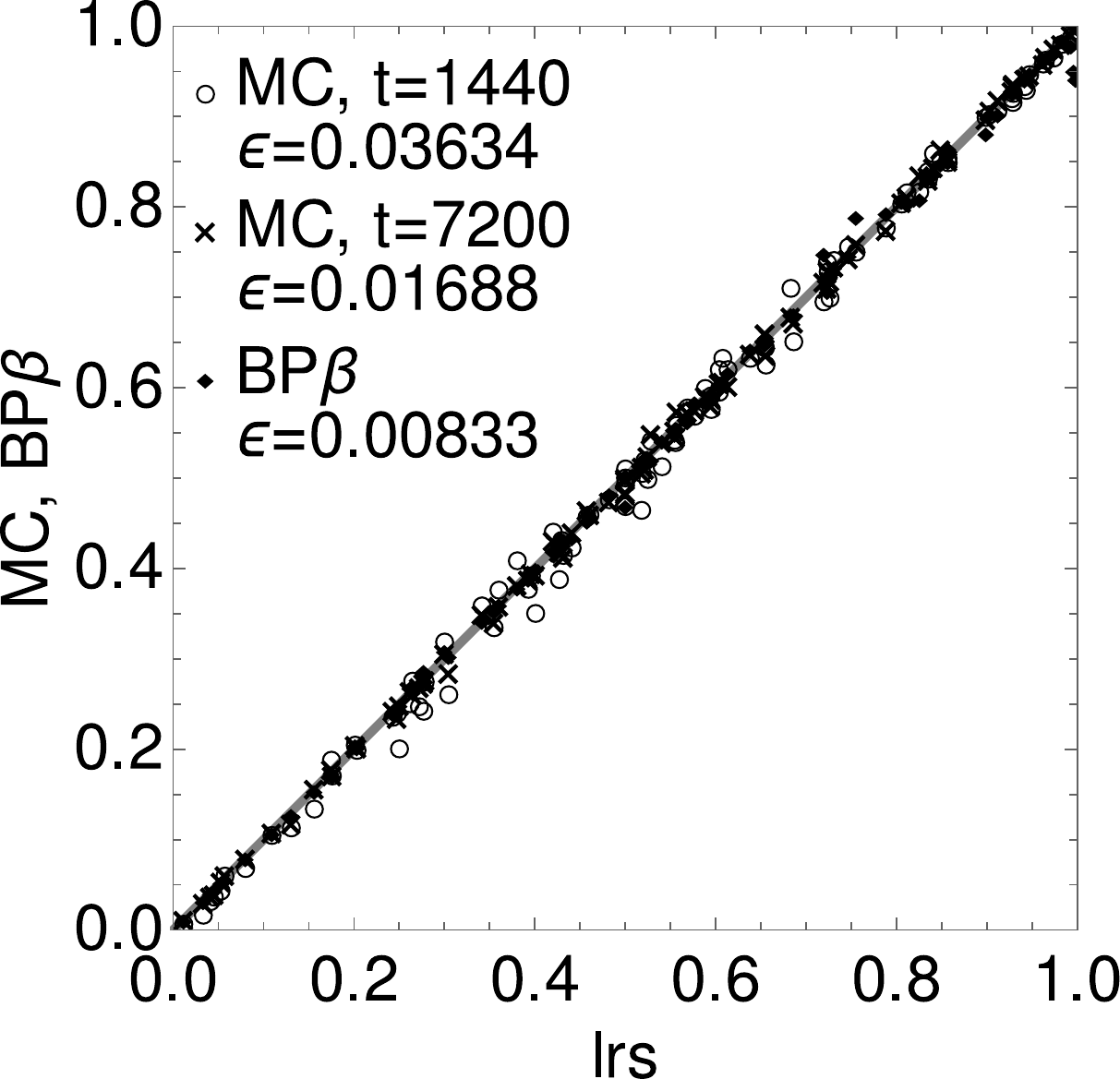}
		\caption{}\label{fig:BPvslrs}
	\end{subfigure}	
	\begin{subfigure}{.45\textwidth}
		\centering
		\includegraphics[width=\linewidth]{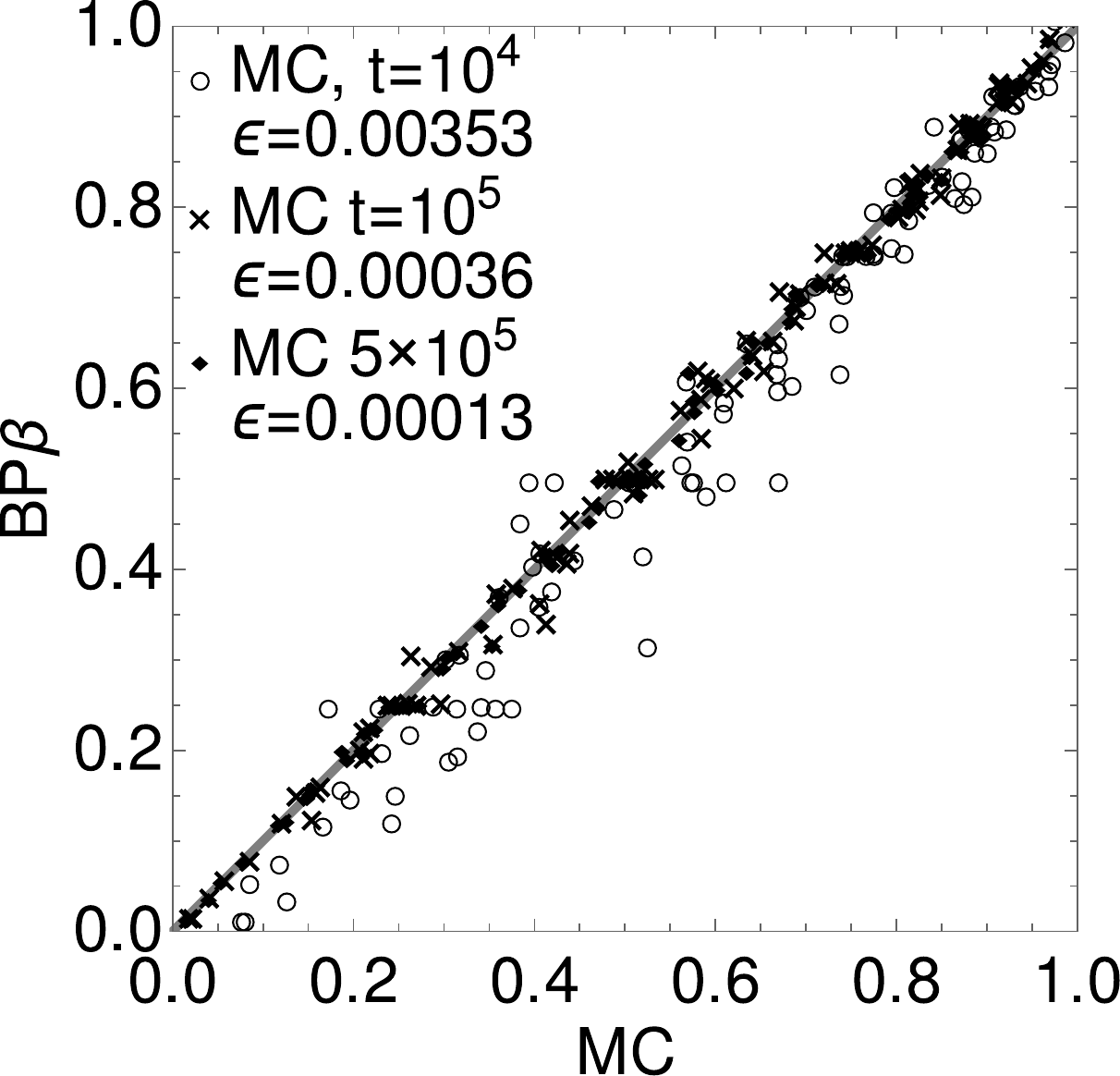}
		\caption{}\label{fig:BPvsMC}
	\end{subfigure}
	\caption{Comparison of algorithms for the computation of the volume of the solution space of \eqref{eq:matrix}. Panel \subref{fig:BPvslrs} compares the exact volume computations given by {\em lrs} with our algorithm (BP-$\beta$) and Monte Carlo with two sampling times: Cobra default (sample size scales  linearly with system size) and a quadratic sample (sampling time scales quadratically with system size). Each point is a random realization of \eqref{eq:matrix} with 12 variables and 4 linear equations involving 3 variables each.  A comparison of Monte Carlo and BP-$\beta$ for a larger system (100 variables, 25 linear equations, 5 variables per equation average) is shown in panel \subref{fig:BPvsMC}. Three sampling times are shown for Monte Carlo: $t_{\text{MC},1}=10^4$, $t_{\text{MC},2}=10^5$ and $t_{\text{MC},3}=5\times10^5$. As the sample time increases, the results agree more with those given by BP$\beta$, indicating that BP$\beta$ is more accurate than Monte Carlo with these sample times. In both panels $\epsilon$ is the mean relative error of the method on the $y$-axis with respect to method on the $x$-axis. ({\em i.e.}, $\epsilon = \frac{1}{Q} \sum_{i=1}^Q \frac{|V_x^i-V_y^i|}{V_x^i}$, where $Q$ is the number of random networks studied).}
\end{figure}

We start this section comparing the output of our  algorithm with the results of exact techniques in small instances of the problem. This is shown in Figure \ref{fig:BPvslrs} where we present a comparison (black rhombuses) between the volumes estimated using our implementation of BP$\beta$ with the results obtained with {\em lrs} \cite{avis1990pivoting_B-237}, an exact scheme to compute the volume of low dimensional polytopes. In the data shown, we worked with random sparse adjacency matrices ${\bf S}$ of dimensions $N=12$ and $M=3$, with an average of $\langle k\rangle=5$ variables participating in each equation. All the non-null elements of matrices were extracted from the set $\{1,-1\}$ with equal probability but guaranteeing, to avoid inconsistencies,  that each equation contains both elements of the set. All the variables were constrained to the interval $[0,1]$.

To have a reference about the quality of the algorithm we present in the same figure results obtained with {\em hit and run} Monte Carlo (abbreviated to MC from now on) simulations. Since it is known that the mixing time of MC (the sample size to guarantee statistical independence) scales quadratically with the system size \cite{demartino2014uniform_1312}, we used a sampling time quadratic in $N$ for all our simulations involving MC. In Figure \ref{fig:BPvslrs} two sampling times were used: $t_{\text{MC},1}=10\times N^2=1440$ (white circles) and $t_{\text{MC},2}=50\times N^2=7200$ (crosses). As can be clearly observed the dispersion of results from MC simulations decreases as $t_\text{MC}$ increases, becoming similar to the dispersion of BP$\beta$. A more quantitative assessment of the dispersion may be obtained by using, as a figure of merit, the average relative error between distinct methods and {\em lrs} ({\em i.e.},
$\epsilon_\text{method} = \frac{1}{Q} \sum_{i=1}^Q \frac{|V_\text{method}^i-V_\text{lrs}^i|}{V_\text{lrs}^i}$, where $Q$ is the number of samples studied, and by `method' we mean: MC at both sample times and BP$\beta$). The corresponding errors are $\epsilon_{\text{MC},1}=0.0363$, $\epsilon_{\text{MC},2}=0.0168$ and $\epsilon_{\text{BP}\beta}=0.0083$, indicating that our algorithm provides results consistent with long runs of {\em hit and run} MC.

In Figure  \ref{fig:BPvsMC} we checked the differences between the volumes computed using BP$\beta$ and MC simulations for larger networks, where exact computations using {\em lrs} are not feasible. We consider random systems of size $N=100$, with $M=25$ linear equations and $\langle k\rangle=5$ average variables per equation. We show results for three different MC times, $t_{\text{MC},1}=N^2=10^4$, $t_{\text{MC},2}=10\times N^2 = 10^5$ and $t_{\text{MC},3}=50\times N^2=5\times 10^5$. As can be seen in the figure, the larger the MC time, the lower the dispersion of the data, indicating that BP$\beta$ and MC are converging to the same results. We measured the relative error between BP$\beta$ and MC ({\em i.e.}, $\epsilon = \frac{1}{Q} \sum_{i=1}^Q \frac{|V_\text{MC}^i-V_{\text{BP}\beta}^i|}{V_\text{MC}^i}$). The figure also suggests that if $t_{MC}$ is small, MC underestimates the volumes. Similar results were obtained for other ensembles of adjacency matrices reflecting small world and scale free networks topologies (Supplementary information).

\begin{figure}
	\centering
	\includegraphics[width=\linewidth]{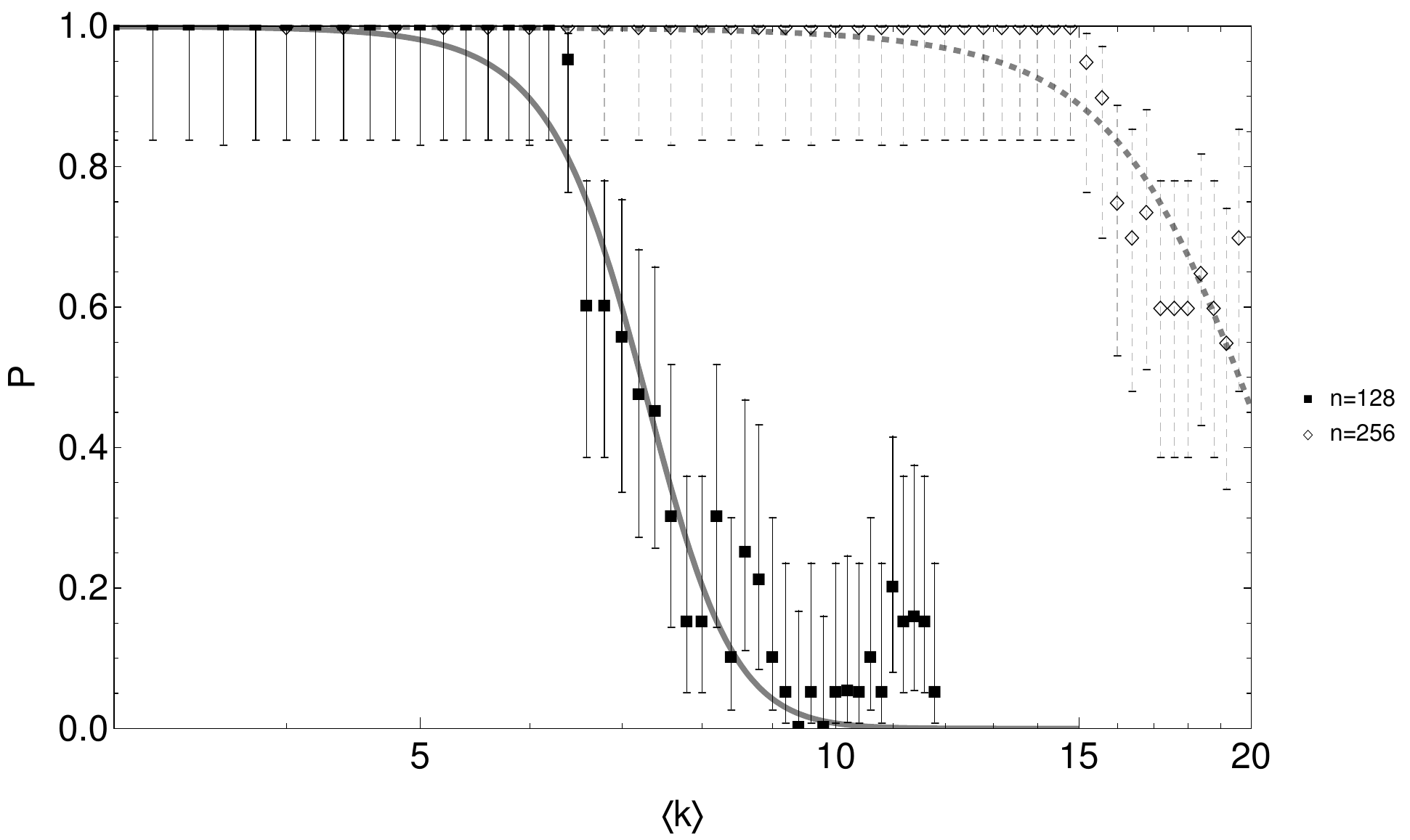}
	\caption{Probability of convergence (P) as a function of the average number of variables per equation ($\langle k \rangle$) in random networks of 64 equations, with 128 and 256 variables respectively. If a network takes more than $10^3$ iterations, we classify it as non-convergence. The fitted curves are intended as visual guides only.}
	\label{fig:convergence}
\end{figure}

Finally, in Figure \ref{fig:convergence} we present the probability of convergence of the algorithm as a function of the number of variables per equation $\langle K\rangle$. In the figure we present data from two ensembles of random matrices, one with $N=128$ (variables), $M=64$ (equations), and the other with $N=256$, $M=64$. In each case, for each value of $\langle k \rangle$ (mean number of variables per equation) we generated 20 random instances and counted the fraction of systems that converged before 1000 iterations. Although this result clearly indicates that our algorithm fails when the matrices involved are too dense, we show below that it may still be useful in many practical situations. Moreover, we see that a larger ratio $N/M$ expands the range of values of $\langle k\rangle$ for which the algorithm converges. 

\subsection{Metabolic Networks}
\label{subsec:metnet}

A metabolic network is an engine that converts metabolites into other metabolites through a series of intra-cellular intermediate steps. The fundamental equation characterizing all functional states of a reconstructed biochemical reaction network is a mass conservation law that imposes simple linear constraints between incoming and outgoing fluxes on every metabolite:
\begin{equation}
\frac{\mathrm{d}\vec{\rho}}{\mathrm{d}t} = {\bf S}\vec{x},
\label{eq:met_dynamic}
\end{equation}
where $\vec{\rho}$ is the vector of the concentrations of the metabolites in the network and $\vec{x}$ is the vector of the reaction fluxes. In this application the matrix ${\bf S}$ (called the stoichiometric matrix) contains the stoichiometric coefficients of each metabolite in each reaction. As long as just steady-state cellular properties are concerned one can assume that a variation in the concentration of metabolites in a cell can be ignored and considered constant. Therefore in case of fixed external conditions one can assume a (quasi) stationarity of the metabolite concentrations and consequently the {\em lhs} of \eqref{eq:met_dynamic} can be set to zero. Moreover, the reaction fluxes are usually restricted by lower and upper bounds defined by biochemical or thermodynamic considerations. Under these very general hypotheses the problem of describing the set of metabolic fluxes compatible with flux-balance constraints \eqref{eq:met_dynamic} is described mathematically by a system of equations identical to \eqref{eq:matrix}.

To validate the applicability of our algorithm in this scenario, we computed the marginal flux distributions of the Red Blood Cell metabolic network, using the stoiciometric matrix presented in \cite{wiback2004monte_437}. The results compare well with the standard Monte Carlo {\em hit and run} sampling (see Supplementary Materials).

We start studying the core of the E.coli metabolism to  understand the effect of enzymopathies in the volume of the solution space of the network. It consists initially of 95 reactions and 72 metabolites \cite{ecolicore}.  The intuition is that the solution space defined by the set of linear equations (\ref{eq:met_dynamic}) and the bounds of the reactions characterizes the size of the phenotypic space of the metabolic network. The larger is the volume of the polytope, the larger is the number of metabolic states of the network.

To reduce the small-loop burden of the network, we removed the smallest molecules. We also eliminated reactions that were inactive for any metabolic state in the medium defined by the exchange fluxes in the network \cite{ecolicore}  (see Supplementary Materials for details). Following \cite{price2004uniform_2172} we reduce to half the bound of one of the reactions while keeping the other bounds fixed, and compute the volume of the modified network. The initial bounds through each reaction are taken from the literature \cite{ecolicore}, to approach as closely as possible a set of metabolic states that actually occur in physiological conditions. The experiment is repeated for every reaction independently.

\begin{figure}
\includegraphics[width=7in]{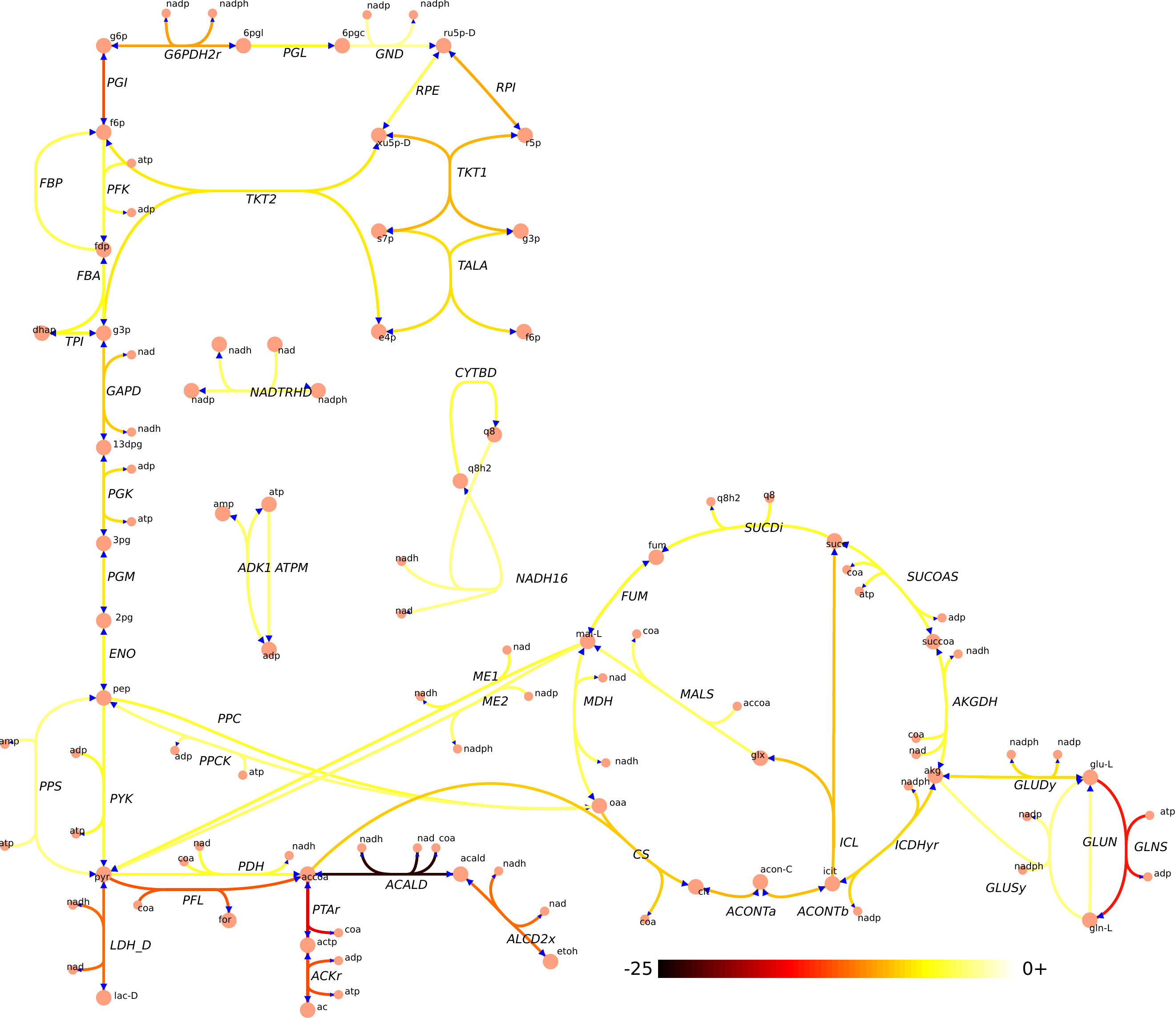}
\caption{Internal reactions of Escherichia coli core metabolism. Each reaction is colored according to the entropy decrease of the space of steady metabolic flux distributions induced by a 50\% flux reduction on that reaction.}
\label{fig:ec_single}
\end{figure}

\begin{figure}
	\includegraphics[width=7in]{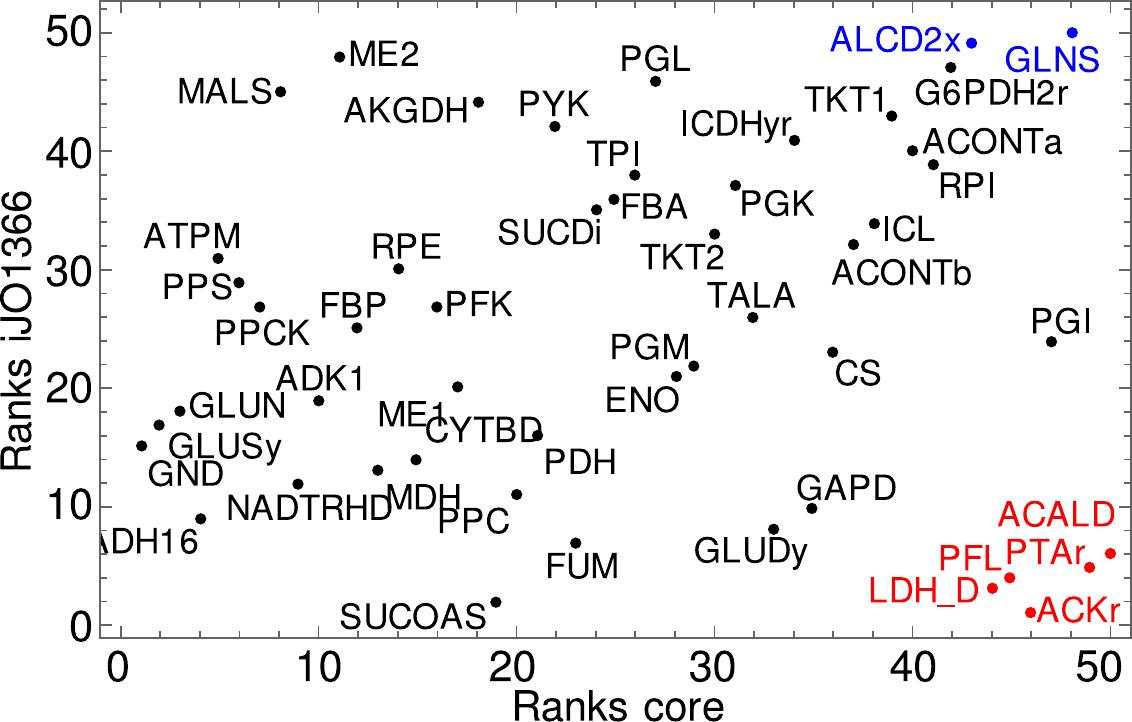}
	\caption{Internal reactions of the core metabolism of E. Coli were sorted according to the impact of a 50\% flux reduction on volume of the solution space of metabolic steady states (high ranking reactions have the largest impacts). We did this for two networks: the E. Coli core metabolic network and the iJO1366 genome-scale reconstruction. Highlighted reactions are discussed in the text: in red a group with markedly different effects in both networks, and in blue reactions with similar impacts in both networks.}
	\label{fig:ec_core_vs_large}
\end{figure}

In Figure \ref{fig:ec_single} we present a  map of the internal reactions of E. coli core metabolic network, where reactions are colored according to the impact on the solution space of the network after a 50\% reduction of their bounds, as explained above. We can observe a highly heterogeneous impact of the distinct reactions. Reactions leading to secretion of lactate or alcohol are very significant in this experiment. Intuitively, this means that blocking the main secretion outlets will severely limit the space of metabolic states available. Also significant are the first reactions of glycolysis and the pentose-phosphate pathways, which are direct destinations of glucose in the core network. The production of glutamine is also important. In the core network glutamine can only be secreted, but this is only because the core network misses most amino-acid metabolism. Another observation is that reactions with lower impacts are those that can be bypassed by alternative pathways. For example, there is only one path to convert glutamate to glutamine, and that is through the reaction GLNS (glutamine synthetase), which has a high impact on the volume. However, there are two alternative routes to the reverse conversion of glutamine to glutamate, through GLUN (glutaminase) and through GLUSy (glutamate synthase), which may explain their smaller impact.

The role of redundancy is more dramatic when we compare these results with a similar experiment on the E. Coli iJO1366 genome-scale metabolic network \cite{ecolilarge}. This large-scale reconstruction has 2583 reactions and 1805 metabolites. We submitted this network to the same reductions as E. Coli core (see Supplementary Materials), leaving a working model with 1417 reactions and 950 metabolites. We performed knock-down of single reactions as described in the previous paragraph for each of the 70 reactions that are also present in the core network. The convergence of the algorithm after a knock-down starting from the wild solution took nearly a minute in a personal computer with an Intel i5-processor. Most of the time of the calculation, approximately 10 minutes per knock-down, was spent calculating the volume.

Figure \ref{fig:ec_core_vs_large} shows a comparison of the ranking of the 70 core reactions according to the impact of the KO on the core network and on the genome-scale iJO1366 network. Interestingly, many reactions have different effects on the genome-scale network than they did in the core network. For the most part this is explained by the higher redundance of iJO1366. Some reactions that have a great impact on the core network, are not significant on the genome-scale network, such as ACALD (acetaldehyde dehydrogenase), PTAr (phosphotransacetylase), PFL (pyruvate formate lyase), LDH (lactate dehydrogenase), ACKr (acetate kinase). In the core network these reactions are directly associated with the production of secreted metabolites: ethanol, acetate, lactate, and, more importantly, cannot be bypassed by alternative pathways. In iJO1366, ethanol and lactate can be produced by other reactions. For example, in iJO1366, lactate is also a product of GLYOX (hydroxyacylglutathione hydrolase) and ACM6PH (N-acetylmuramate 6-phosphate hydrolase), neither of which are present in the core network. Another example is ACALD, which in the core network is the sole producer of acetaldehyde, which is then converted to ethanol. In iJO1366, acetaldehyde can also be produced by DRPA (deoxyribose-phosphate aldolase), THRAi (threonine aldolase) and ETHAAL (Ethanolamine ammonia-lyase), among others. This warns us that while the core network captures many essential features of E. Coli metabolism, a full account of redundancy capabilities is only found on the genome-scale model.

On the other hand, some reactions are seen to be very important in both networks. These are the reactions which are not bypassed by alternative pathways in the iJO1366 reconstruction. We highlight GLNS and ALCD2x. The core network lacks amino-acid metabolism, so glutamine is secreted or directly involved in biomass production; its role here is that of an outlet. In the genome-scale model, glutamine opens the door to a network of reactions that constitute amino-acid metabolism. So, for different reasons, GLNS is significant in both models. On the other hand, ALCD2x (alcohol dehydrogenase) is the sole responsible for the production of ethanol in both core and iJO1366 networks, obtaining it from the conversion of acetaldehyde. The lack of an alternative pathway for ethanol production bypassing ALCD2x justifies its simultaneous importance in both reconstructions.

\subsection{Network Tomography}
\label{subsec:nettom}

As a second application example of our algorithm we present a problem from the field of Network Tomography. A communication network is a collection of nodes representing computer terminals, routers, or subnetworks. Two nodes are linked if there is a direct connection between them that does not involve other nodes. Messages are transmitted by sending {\em packets} of bits from a {\em source} node to a {\em destination} node along a path consisting of one or more links and which generally passes through other nodes. In many real world applications it is not feasible to keep a centralized record of the origins and destinations of all the packets that have traversed the network, either because of bandwidth restraints or hardware limitations. The count of packets traversing each link is a more readily available datum. In this section we show that the BP$\beta$ algorithm can be used to estimate the traffic flow between all source-destination pairs, given measurements of the traffic flow in each link.

We consider fixed routing networks, where the path between all source-destination pairs is known and remains the same for all packets traveling between these two nodes. For simplicity of notation source-destination pairs will be identified with a single letter index. Thus, let $x_a(t)$ be the traffic intensity (packets per unit time) on the $a$'th source-destination pair at time $t$. The total traffic through the $i$'th link, which we denote as $y_i(t)$, is given by the sum of the source-destination flows where the path traverses the $i$'th link. If we define a routing matrix $\mathbf{S}$ with components $S_{ai}=1$ if the $i$'th link participates in the path associated to the $a$'th source-destination pair, and $S_{ai}=0$ otherwise, it follows that
\begin{equation}
\vec{y}(t)=\mathbf{S}\vec{x}(t).
\label{eq:abilene}
\end{equation} 
Then, given measurements of link traffic $\vec{y}(t)$, the problem is to estimate $\vec{x}(t)$,  subject to the non-negativity constraint $\vec{x}(t)\ge0$. Equation \eqref{eq:abilene} is of the same form as \eqref{eq:matrix}. To make an analogy with metabolic networks, the metabolites are mapped in this context to the links of the communication network, and the reactions become source-destination pairs.

\begin{figure}
\includegraphics[width=4in]{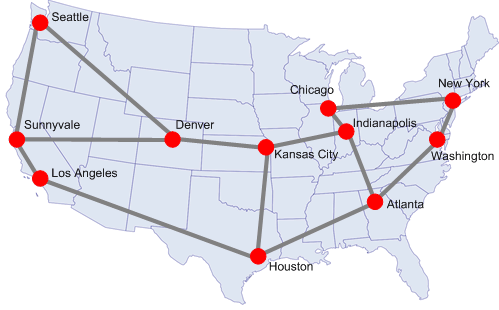}
\caption{Abilene internet network.}
\label{fig:abilene}
\end{figure}

\begin{figure}
\centering
	\begin{subfigure}{.35\textwidth}
		\centering
		\includegraphics[width=\linewidth]{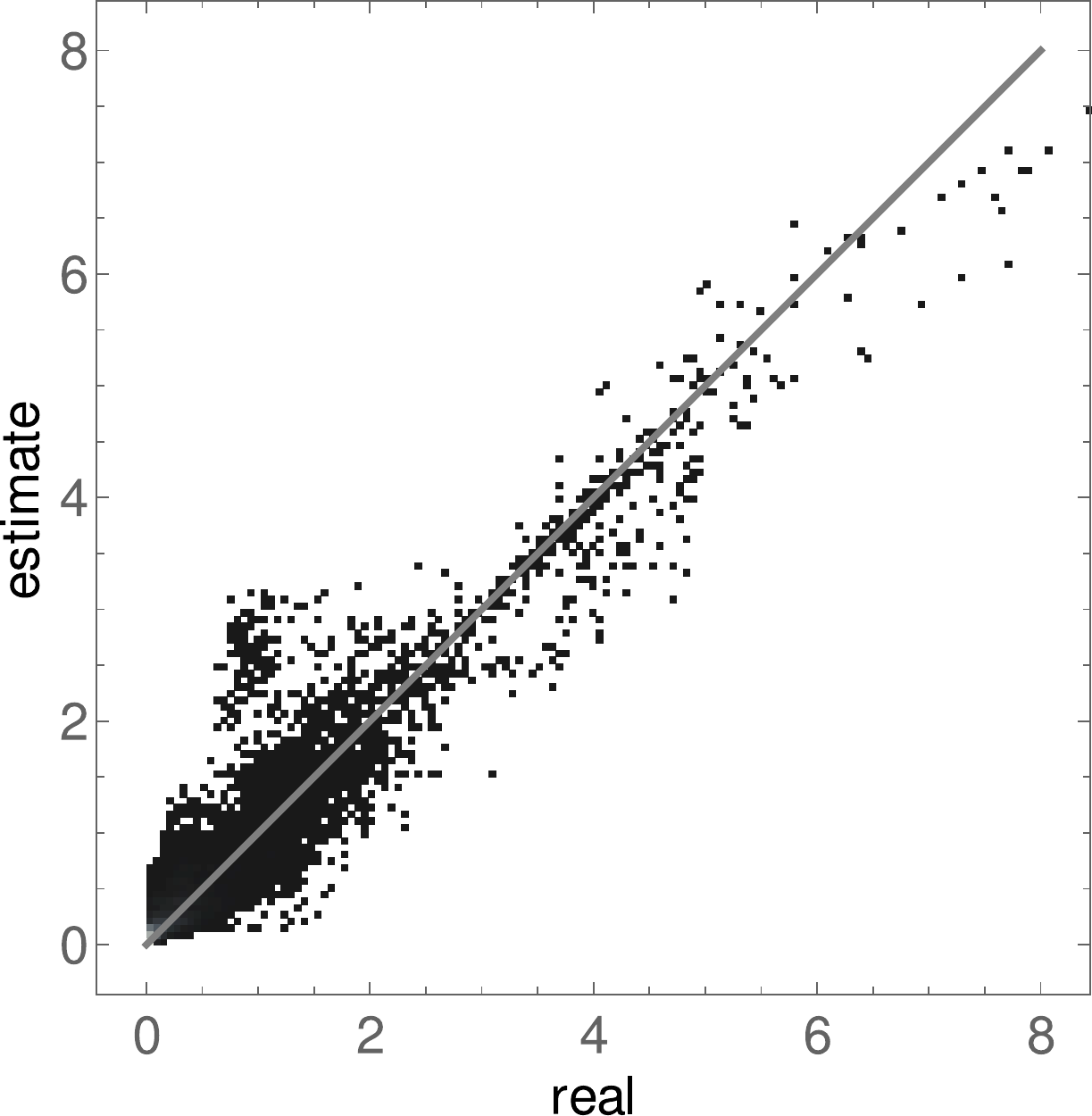}
		\caption{}\label{fig:abilene_scatter}
	\end{subfigure}	
	\begin{subfigure}{.52\textwidth}
		\centering
		\includegraphics[width=\linewidth]{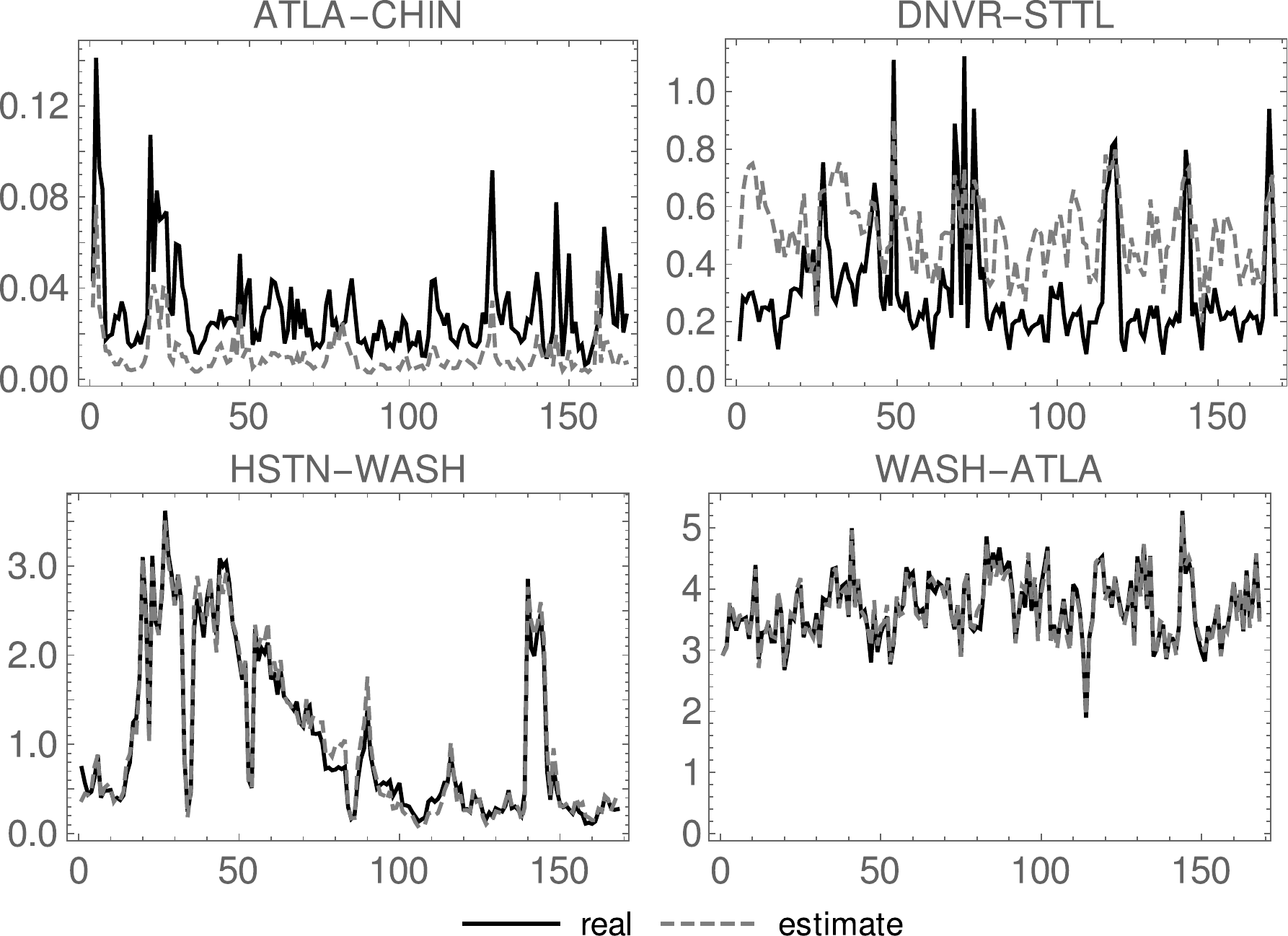}
		\caption{}\label{fig:abilene_time}
	\end{subfigure}
	\caption{Panel \subref{fig:abilene_scatter} shows the correlation between estimated and real flows, aggregated over all origin-destination pairs and all times. Panel \subref{fig:abilene_time} shows a few representative origin-destination flow estimates (discontinuous, gray) compared to real measurements (black) as a function of time. The average relative error over all origin-destination pairs and all times is 0.3.}
\end{figure}

As a working example we use real world data from a portion of the Internet, a sub-network known as Abilene \cite{kolaczyk2009statistical}. Measurements of origin-destination flows were taken continuously over a seven-day period, starting on December 22, 2003. The relevant portion of the Abilene network, as it was at that time, may be conceptualized as a graph consisting of 11 nodes and 30 links (see Figure \ref{fig:abilene}). The original traffic flow counts were aggregated to five-minute intervals, a standard approach used to avoid issues of time synchronization across the network. We further averaged to 1-hour intervals to mitigate the effects of noise \cite{zhang2003fast_206}. From the origin-destination traffic data, we generated link flows using Equation \eqref{eq:abilene}. Then from the link flow data, we try to estimate the original origin-destination traffic. These estimates are calculated as averages over the marginal distributions returned by BP-$\beta$ using as upper bounds the maximum flow observed throughout the experiment.

Figure \ref{fig:abilene_scatter} shows a scatter plot of the aggregated traffic over all origin-destination pairs and for all times. It is clear from the figure that there is a linear correlation between real and inferred flows, with an aggregated root mean square error of 0.07. A peculiar cluster of points in the lower-left corner shows small but systematic overestimation. Closer inspection reveals that these points originate from a single origin-destination pair in the network: Denver-Indianapolis. The path associated with this pair consists of the single link Denver-Indianapolis, which participates in over 15 other origin-destination paths in the network. Thus the traffic measurements on this link give very poor information about the flow on the Denver-Indianapolis path. Under these circumstances inference about the path Denver-Indianapolis is hard, and our algorithm is doing the best it can with the information given. Any improvement on this should come from extra-information about the network.

To have a more intuitive picture of the quality of our algorithm we show in Figure \ref{fig:abilene_time} the estimates and the experimental traffic as functions of time for four different origin-destination links. The relative error of our estimations, considering all origin-destination pairs is 0.3. More sophisticated algorithms, which make assumptions about time correlations or incorporate prior probabilities, score average relative errors between 0.1 and 0.4 \cite{kolaczyk2009statistical, zhang2003information_301, lakhina2003structural_}. Our simple method highlights the value of inferences drawn directly from the structure of the problem without incorporating {\em a priori} assumptions.

\section{Conclusions}

In this work we presented an efficient and robust algorithm to estimate the volume of a convex polytope in high dimensions. The algorithm allows the direct computation of the marginal distributions of variables defined as the solution of linear ill-posed problems in a convex space. Our implementation provides results that are statistically indistinguishable from other numerical techniques, with much less computational effort, it scales linearly with the system size, and has a finite operation cost per iteration. The algorithm was tested in various applications, starting with random systems of equations to show its robustness. Then, taking profit of its efficiency, we studied an E. Coli genome scale metabolic network for multiple knock-downs, unveiling the importance of non-redundant reactions in the functional flexibility of the network. We compare with the effects of knock-downs in the core metabolic network of E. Coli, obtaining significant differences. This highlights the abundant redundancy in the genome-scale model that is not present in the core model. On the other hand, knock-down of some reactions have similar effects in both networks, for example alcohol production, since its important role as a secreted metabolite is captured in both models. Differences in other reactions, such as ACALD, which is essential in the core network for the production of ethanol, can be explained by the presence of alternative pathways to the same products in the genome-scale model. As a final working example we studied a problem from the field of Network Tomography, the estimation of the origin-destination traffic in a communication network, from the traffic flow in the indiviudal links. Again, our results compare statistically well with other numerical estimates that make assumptions about correlations or incorporate prior probabilities about the traffic flow. As a by-product we also showed that, for specific origin-destination pairs, in particular the Denver-Indianapolis pair of the Abilene network, the correct estimation of the corresponding origin-destination flow cannot be done without introducing additional external information.

\bibliographystyle{unsrt}
\bibliography{polytope}

\clearpage

\appendix
\section{Supplementary Material}

\subsection{Methods and Algorithms}

We propagate messages through the rules \eqref{eq:update_ai} and \eqref{eq:update_ia}, where the proportionality constants ensure that the messages stay normalized. The marginals $\mathrm{P}_a(\{x_j\}_{j \in a})$ and $\mathrm{P}_i(x_i)$ (Equation \eqref{eq:Pa} and \eqref{eq:Pi}) are estimated as:
\begin{equation}
\mathrm{P}_a(\{x_j\}_{j \in a}) = Z_a^{-1} 
	\delta\left(\sum_{j\in a} S_{aj}x_j\right)
	\prod_{i\in a} m_{i\rightarrow a}(x_i),
\quad
\mathrm{P}_i(x_i) = Z_i^{-1} \prod_{a\in i} m_{a\rightarrow i}(x_i),
\label{eq:marginals}
\end{equation}
where $Z_a$ and $Z_i$ are normalization constants. We will parametrize the messages as truncated generalized Beta distributions:
\begin{equation}
m(x) = Z^{-1}(x-A)^{\alpha-1}(B-x)^{\beta-1},
\label{eq:message_parameterization}
\end{equation}
if $A<x<B$ and $m(x)=0$ otherwise, where $Z=(B-A)^{\alpha+\beta-1}\mathrm{B}(\alpha,\beta)$ is a normalization constant. Our job now is to reformulate the update rules \eqref{eq:update_ai} and \eqref{eq:update_ia} in terms of the message parameterization \eqref{eq:message_parameterization}. But first we need some definitions. Let $\mu$, $\sigma^2$ denote the mean and variance of the generalized Beta distributions:
\begin{equation}
\mu=\frac{A\beta+B\alpha}{\alpha+\beta}, \quad
\sigma^{2}=\frac{\alpha\beta(B-A)^2}{(\alpha+\beta)^2 (1+\alpha+\beta)}.
\label{eq:cum_from_alpha_beta}
\end{equation}
These relations can be inverted to give $\alpha,\beta$ in terms of $\mu,\sigma^2$:
\begin{equation}
\alpha=\lambda\frac{\mu-A}{B-A}, \quad
\beta=\lambda\frac{B-\mu}{B-A}
\label{eq:alpha_beta_from_cum}
\end{equation}
where $\lambda=(\mu-A)(B-\mu)/\sigma^2-1$.

We begin with the update rule for $m_{a\rightarrow i}$. Equation \eqref{eq:update_ai} says that $x_i$ in $m_{a\rightarrow i}$ distributes as the sum $-S_{ai}^{-1}\sum_{j\in a\backslash i} S_{aj}x_j$, where $x_j\sim n_{j\rightarrow a}$ and $A_{j\rightarrow a}\le x_j\le B_{j\rightarrow a}$. We have:
\begin{equation}\begin{aligned}
A_{a\rightarrow i} & =
	\min_{\{x_j\}_{j\in a\backslash i}}
		\left(- \sum_{j \in a \backslash i} \frac{S_{aj}}{S_{ai}} x_j\right) =
	- \sum_{j\in a^- \backslash i} \frac{S_{aj}}{S_{ai}} A_{j \rightarrow a}
	- \sum_{j\in a^+ \backslash i} \frac{S_{aj}}{S_{ai}} B_{j \rightarrow a}
\\
B_{a\rightarrow i} & =
	\max_{\{x_j\}_{j\in a\backslash i}}
		\left(- \sum_{j \in a\backslash i} \frac{S_{aj}}{S_{ai}} x_j\right) =
	- \sum_{j\in a^- \backslash i} \frac{S_{aj}}{S_{ai}} B_{j \rightarrow a}
	- \sum_{j\in a^+ \backslash i} \frac{S_{aj}}{S_{ai}} A_{j \rightarrow a}
\end{aligned}
\label{eq:ABupdate}
\end{equation}
where $j\in a^+$ denotes the reactions $j$ for which $\slfrac{S_{aj}}{S_{ai}}>0$, and $j\in a^-$ denotes those for which $\slfrac{S_{aj}}{S_{ai}}<0$. Since the mean and variance are additive under convolutions,
\begin{equation}
\mu_{a\rightarrow i} = 
	- \sum_{j\in a\backslash i} \frac{S_{aj}}{S_{ai}} \nu_{j\rightarrow a},\quad
\sigma_{a\rightarrow i}^2 = 
	\sum_{j\in a\backslash i} \frac{S_{aj}^2}{S_{ai}^2} \tau_{j\rightarrow a}^2.
\end{equation}
From these we can calculate $\alpha_{a\rightarrow i}$, $\beta_{a\rightarrow i}$ using \eqref{eq:alpha_beta_from_cum}. If $A_{a\rightarrow i}< a_i$ or $B_{a\rightarrow i} > b_i$ in \eqref{eq:ABupdate}, we find the mean and variance of the Beta distribution truncated to the interval $[\max(A_{a\rightarrow i}, a_i), \min(B_{a\rightarrow i}, b_i)]$, and from these recompute the values $\alpha_{a\rightarrow i}$, $\beta_{a\rightarrow i}$, and reset $A_{a\rightarrow i} := \max(A_{a\rightarrow i}, a_i)$, $B_{a\rightarrow i} := \min(B_{a\rightarrow i}, b_i)$. This way we guarantee that message $m_{a\rightarrow i}$ always has its support contained in the interval $[a_i,b_i]$. This completes the updating of $m_{a\rightarrow i}$. It should be noted that the convolution of two Beta distributions is not an exact Beta distribution, however this approximation behaves well in practice \cite{johannesson1995approximations_489}. 

Now we analyze the update rule for $m_{i\rightarrow a}$. From \eqref{eq:update_ia} we have that:
\begin{equation}
A_{i\rightarrow a} = \max_{c\in i\backslash a}(A_{c\rightarrow i}),\quad B_{i\rightarrow a} = \min_{c\in i\backslash a}(B_{c\rightarrow i}).
\end{equation}
Next, using widely available efficient single-variable numerical integration routines \cite{gsl}, we can calculate:
\begin{equation}
\mu_{i\rightarrow a} = 
\frac{\int_{A_{i\rightarrow a}}^{B_{i\rightarrow a}} x \prod_{c\in i\backslash a} m_{c\rightarrow i}(x) \mathrm{d}x}{\int_{A_{i\rightarrow a}}^{B_{i\rightarrow a}} \prod_{c\in i\backslash a} m_{c\rightarrow i}(x) \mathrm{d}x}, \quad
\sigma_{i\rightarrow a}^2 = 
\frac{\int_{A_{i\rightarrow a}}^{B_{i\rightarrow a}} x^2 \prod_{c\in i\backslash a}m_{c\rightarrow i}(x) \mathrm{d}x}{{\int_{A_{i\rightarrow a}}^{B_{i\rightarrow a}} \prod_{c\in i\backslash a} m_{c\rightarrow i}(x) \mathrm{d}x}} - \mu_{i\rightarrow a}^2
\end{equation}
and from these we calculate $\alpha_{i\rightarrow a}$ and $\beta_{i\rightarrow a}$ using \eqref{eq:alpha_beta_from_cum}. This completes the updating of $m_{i\rightarrow a}$.

The entropy of the system is computed using numerical integration routines from GSL \cite{gsl}. In particular, the first term in \eqref{eq:Sa} is a convolution, and can be readily computed using the Fourier transform of the Beta distribution, which is a confluent hypergeometric function of the first kind.

\subsection{Red Blood Cell}

\begin{figure}
	\includegraphics[height=3in,width=5in]{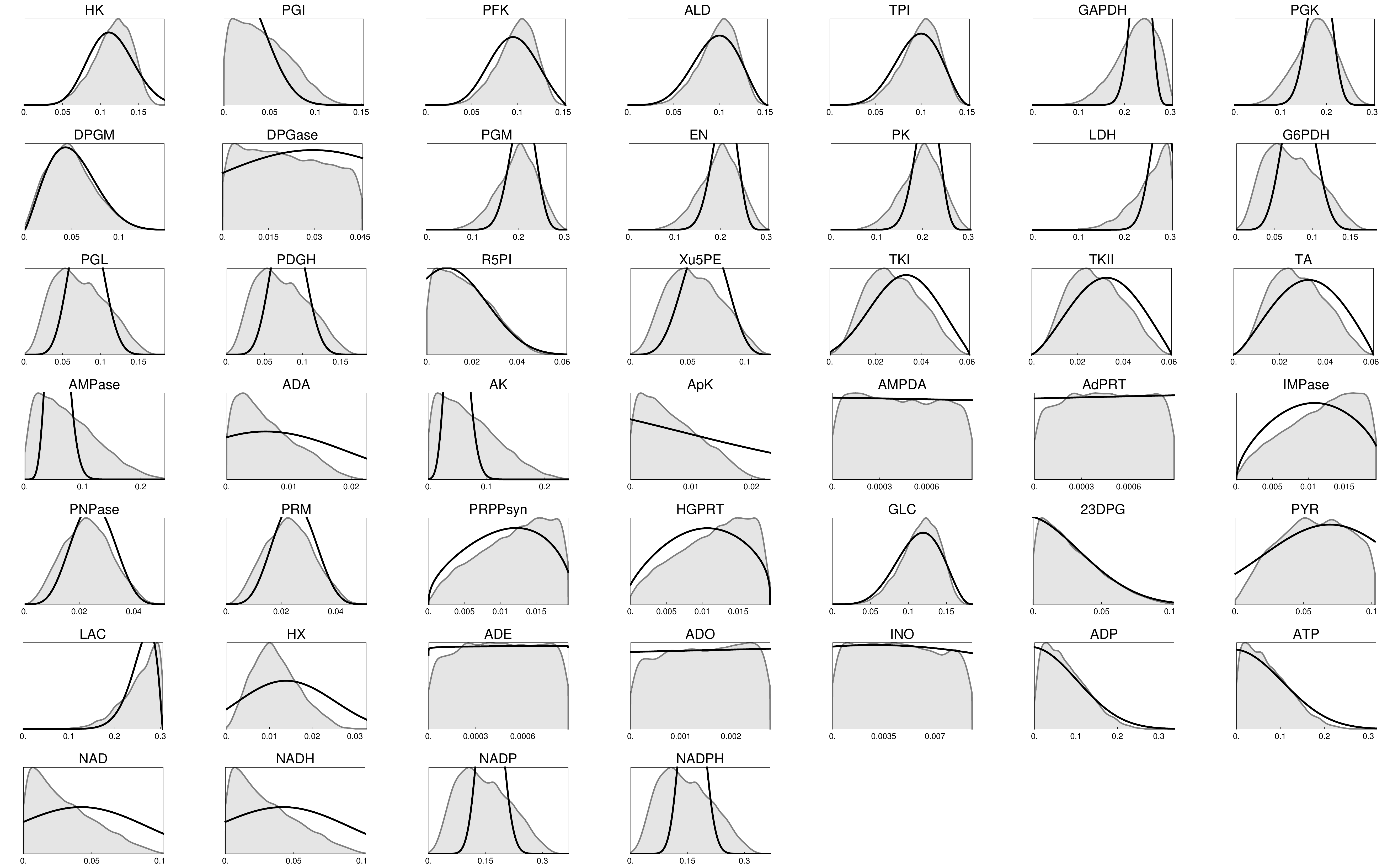}
	\caption{Flux distributions of the reactions in the Red Blood Cell metabolic network, computed with Monte Carlo {\em hit and run} (gray, filled) and with our BP-$\beta$ algorithm (black contours).}
	\label{fig:rbc}
\end{figure}

We used BP$\beta$ algorithm and Monte Carlo {\em hit and run} sampling to obtain the marginal flux distributions for each of the reactions in the Red Blood Cell metabolism taking the same stoichiometric matrix presented in \cite{wiback2004monte_437}. The network contains 46 reactions and 34 metabolites. All reactions are irreversible, with upper bounds set to realistic physiological values given in \cite{wiback2004monte_437}. We compared BP$\beta$ with a set of 10000 feasible solutions generated by Monte Carlo sampling. As seen in Figure \ref{fig:rbc}, the predictions of both methods compare rather well.

\subsubsection{Modifications in E. Coli metabolic network}

To reduce the small-cycle burden of the network, we eliminated the smallest molecules from the list of metabolites. By this criteria we manually eliminated the following metabolites: o2, h2o, nh4, pi, h, co2. We also deleted reactions that were constrained to zero-flux by the nutrients available, as defined by the exchange fluxes in the network (reaction bounds were taken from \cite{ecolicore}). After these two steps, we deleted metabolites that did not participate in any reaction, as well as empty reactions.

We also removed the reaction of biomass production, since it was not our interest to bias the solution space with this objective function. In order to model the net production of metabolites in the network, which in physiological conditions are diverted to various activities in the cell, such as biomass production, protein synthesis, or even degradation, we included drain reactions for every metabolite. Finally, pairs of irreversible reactions that were mirrors of each other were merged into a single reversible reaction.


\subsection{Scale free and Small-world Random Networks}

\begin{figure}
\centering
	\begin{subfigure}[t]{.45\textwidth}
		\centering
		\includegraphics[width=\textwidth]{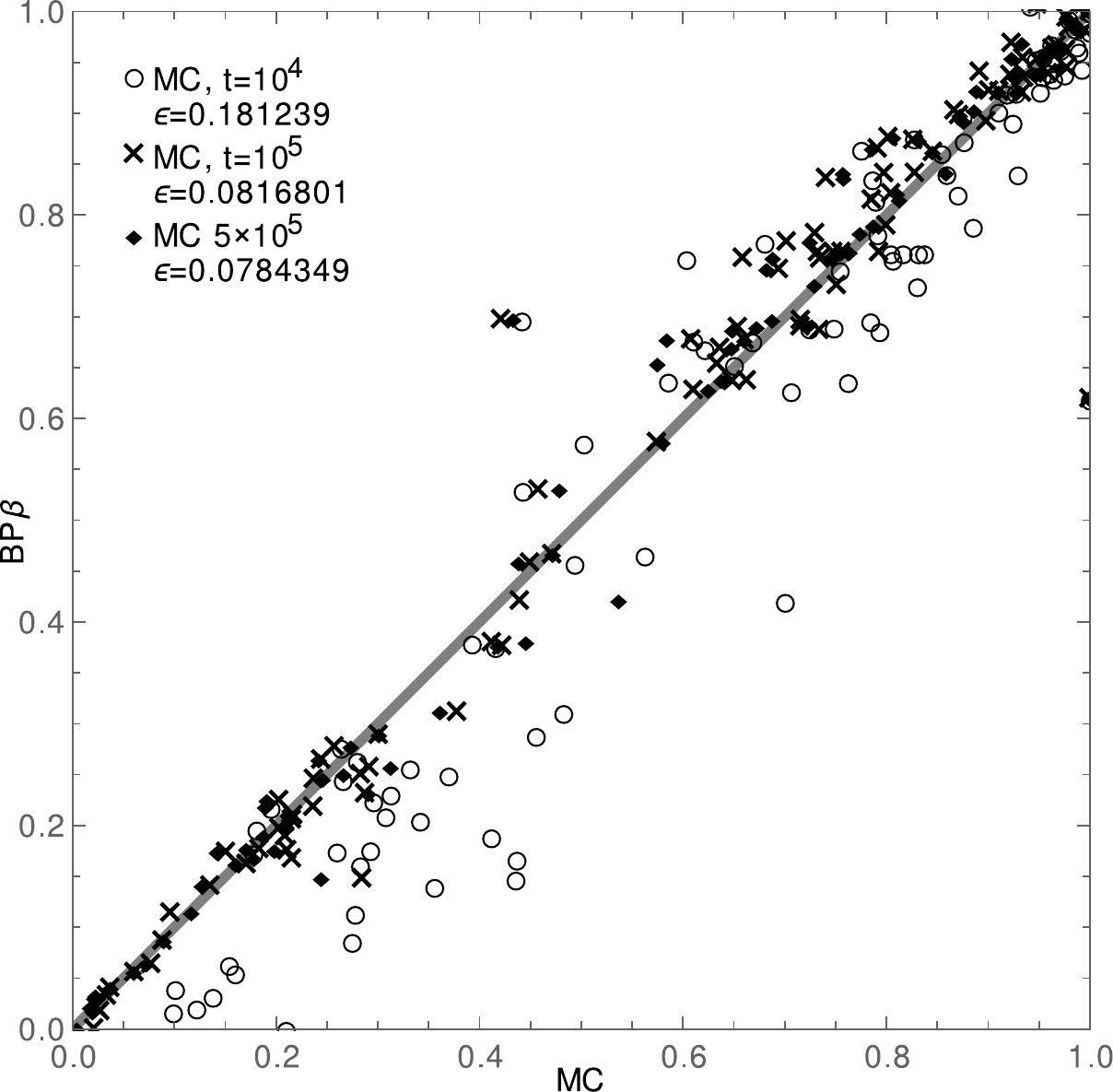}
		\caption{}\label{fig:BPvsMCsw}
	\end{subfigure} %
	\hfill
	\begin{subfigure}[t]{.45\textwidth}
		\centering
		\includegraphics[width=\textwidth]{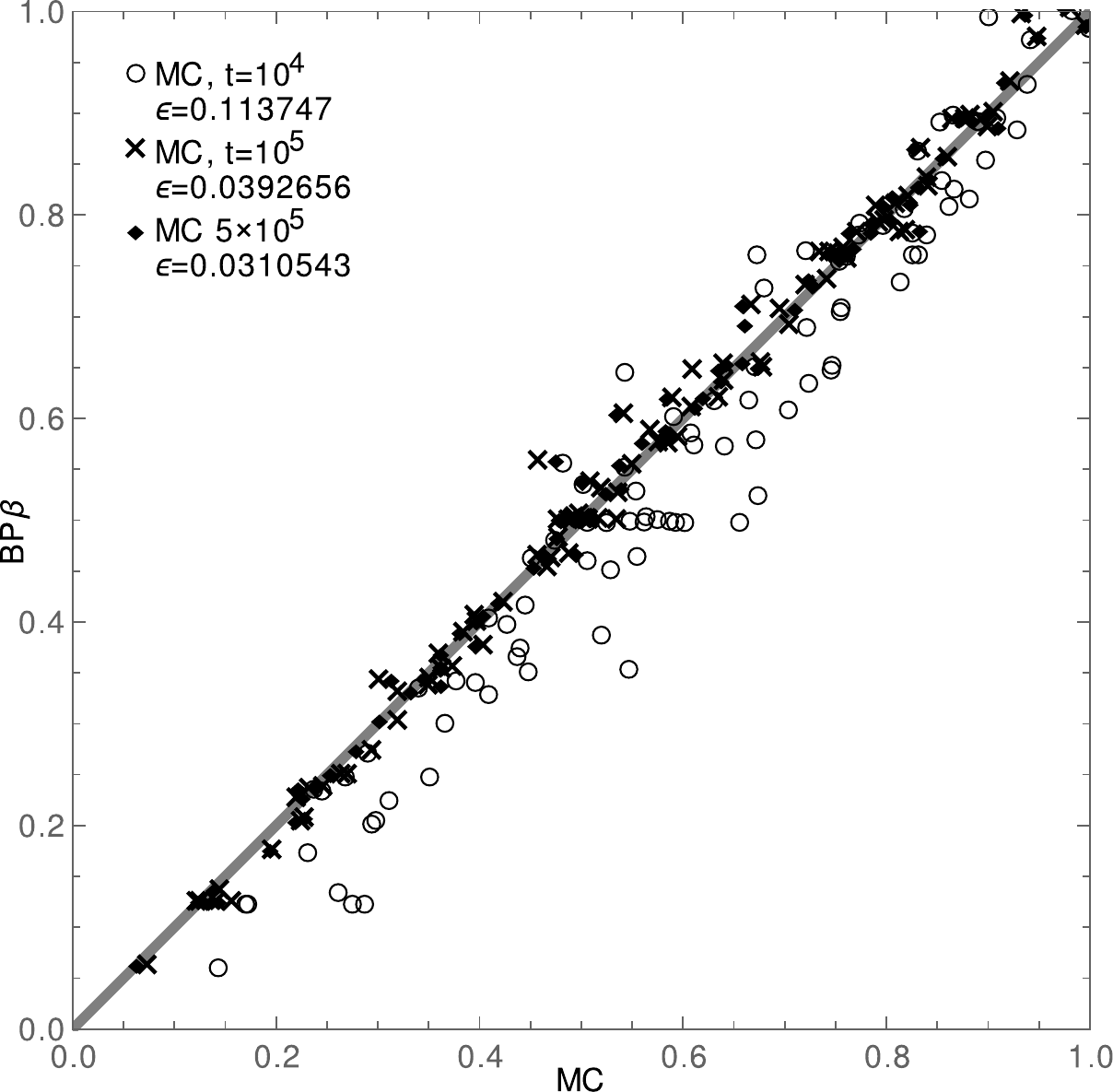}
		\caption{}\label{fig:BPvsMCsf}
	\end{subfigure}
	
	\bigskip
	
	\caption{Comparison of algorithms for the computation of the volume of the solution space of \eqref{eq:matrix}, as in \ref{fig:BPvsMC}, but using two distinct random network models. Panel \subref{fig:BPvsMCsw} does the comparison on a set of random networks generated according to the small world model. Panel \subref{fig:BPvsMCsf} does the comparison on a set of random networks generated according to the scale free model.}
\end{figure}

Figure \ref{fig:BPvsMC} compares the accuracy of our algorithm with that of a Monte Carlo {\em hit and run} method on a set of random networks generated according to the \"Erdos \& R\'enyi model \cite{erdoes1960evolution_17}. We also did this comparison on two additional models of random networks: the small-world model and the scale free model \cite{reka2002statistical_47}. As can be seen in Figures \ref{fig:BPvsMCsw} and \ref{fig:BPvsMCsf}, the results are qualitatively the same as in \ref{fig:BPvsMC}, validating the applicability of the algorithm on real networks.

In the small world model the random construction begins with a network embedded in space where each node is linked to some of its closest neighbors. Then random links are inserted between random pairs of nodes independently of the distance between them, bringing closer together nodes that are far apart in the original spatial arrangement of the network (hence the name ``small world''). In the scale-free model nodes are linked randomly with a preference for attachment to nodes that already have the largest degrees. The small-world and the scale-free models of random networks were created to try to explain properties observed in real large-scale networks that weren't present in the original random network model of \"Erdos \& R\'enyi \cite{reka2002statistical_47}.

To generate small-world random systems of equations, we started with a factor graph consisting of a cycle of alternating equations and variables. Then we added links connecting random pairs of equations and variables. To generate scale-free random systems of equations, we started with a network where each equation is connected to two variables with opposite signs. This ensures that the network is consistent. Then we added random variables, which have a probability of linking an equation proportional to that equation's degree. Each added variable has a fixed degree of 3.

\subsection{Supplementary programs}

The C++ source code of the BP$\beta$ algorithm used in the simulations in this paper are available from the authors upon request.

\end{document}